\documentclass[a4paper,10pt]{article}
\usepackage[cp1251]{inputenc}
\usepackage[english]{babel}
\usepackage{amsmath}
\usepackage{amssymb}
\usepackage{graphicx}
\usepackage[pdftex]{color}
\usepackage[sort&compress,square,numbers,comma]{natbib}
\usepackage[pdftex]{hyperref}
\hypersetup{
             final,
             colorlinks=true,
             linkcolor=blue,
             citecolor=red
}

\begin{document}
\twocolumn[\scriptsize{\slshape 0021-3640, JETP Letters, 2009, Vol. 90, No. 6, pp. 469–474.  \textcopyright\, Pleiades Publishing, Ltd., 2009.}

\scriptsize{\slshape Original Russian Text \textcopyright\, P.V.
Ratnikov, 2009, published in Pis’ma v Zhurnal \'{E}ksperimental’no\u{\i} i Teoretichesko\u{\i} Fiziki, 2009, Vol. 90, No. 6, pp. 515–520.}

\vspace{0.5cm}

\hrule\vspace{0.07cm}

\hrule

\vspace{0.75cm}

\begin{center}
\LARGE{\bf Superlattice Based on Graphene on a Strip Substrate}

\vspace{0.25cm}

\large{\bf P.\,V. Ratnikov}

\vspace{0.1cm}

\normalsize

\textit{Lebedev Physical Institute, Russian Academy of Sciences,}

\textit{Leninski\u{\i} pr. 53, Moscow, 119991 Russia}

\textit{e-mail: \url{ratnikov@lpi.ru}}

Received August 12, 2009
\end{center}

\vspace{0.1cm}
\begin{list}{}
{\rightmargin=1cm \leftmargin=1cm}
\item
\small{A graphene-based superlattice formed due to the periodic modulation of the band gap has been investigated. Such a modulation is possible in graphene deposited on  a strip substrate made of silicon oxide and hexagonal boron nitride. The advantages and some possible problems in the superlattice under consideration are discussed. A model describing such a superlattice is proposed and the dispersion relation between the energy and momentum of carriers has been obtained using the transfer matrix method within this model.}

\vspace{0.05cm}


\vspace{0.05cm}

\small{\bf DOI}: 10.1134/S0021364009180143

\end{list}\vspace{0.75cm}]

\begin{center}
1. INTRODUCTION
\end{center}

Interest in graphene-based superlattices has increa-sed in recent years. Calculations of graphene-based superlattices with periodic rows of vacancies were performed using the molecular dynamics method \cite{Chernozatonskii1}. Calculations of single-atom-thick superlattices formed by lines of pairs of adsorbed hydrogen atoms on graphene were carried out with the density functional theory \cite{Chernozatonskii2}.

Rippled graphene that can be treated as a superlattice with the one-dimensional periodic potential of ripples was investigated in \cite{Isacsson, Guinea,
Wehling}. Superlattices obtained when a periodic electrostatic potential \cite{Bai,
Barbier, Park1, Park2} or periodically located magnetic barriers \cite{Masir1, Masir2, Anna, Ghosh} were applied to graphene were analytically examined.

However, the investigation of the graphene-based superlattice with a periodic electrostatic potential disregarded the fact that the application of the electrostatic potential to a gapless semiconductor (graphene) results in the production of electron–hole pairs and the redistribution of charge s: electrons move from the region where the top of the valence band lies above the Fermi level to the region where the bottom of the conduction band lies below the Fermi level. The superlattice becomes a structure consisting of positively charged regions, where the electrostatic potential displacing the Dirac points upward in energy is applied, alternating with negatively charged regions. The strong electrostatic potential of induced charges appears and strongly distorts the initial step electrostatic potential and, therefore, the electronic structure of the superlattice calculated disregarding the electrostatic potential of induced charges.

To avoid the production of electron–hole pairs, a superlattice appearing due to the periodic modulation of the band gap is considered.

A superlattice in the form of the periodic planar heterostructure of graphene nanoribbons between
which nanoribbons of hexagonal boron nitride (h-BN) are inserted was previously proposed in \cite{Sevincli}. The band structure of such a superlattice was numerically calculated. However, it is very difficult to implement this superlattice even using the advances of modern lithography, because problems inevitably arise with the control of periodicity in the process of the etching of nanoribbons in a graphene sheet and the insertion of h-BN nanoribbons. Moreover, h-BN is an insulator with a band gap of 5.97 eV, which significantly hinders the tunneling of carriers between graphene nanoribbons. Such a heterostructure is most probably a set of quantum wells where the wavefunctions of carriers from neighboring quantum wells almost do not overlap.

\begin{figure}[!b]
\hypertarget{Fig1}{}
\includegraphics[width=0.5\textwidth]{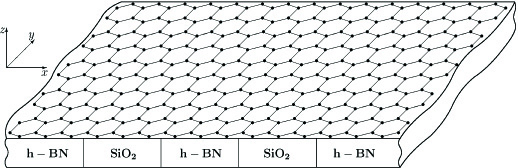}
{\bf Fig. 1.} Graphene on the strip substrate consisting of alternating SiO$_2$ and h-BN strips.
\end{figure}

In this work, a superlattice formed by a graphene sheet deposited on a strip substrate is proposed. The strip substrate is made of periodically alternating strips of SiO$_2$ (or any other material that does not affect the band structure of graphene) and h-BN, as shown in \hyperlink{Fig1}{Fig. 1}. The h-BN layers are located so that its hexagonal crystal lattice is under the hexagonal crystal lattice of graphene. Owing to this location, a band gap of 53~meV appears in the band structure of graphene in the graphene-sheet regions under the h-BN layers \cite{Zhou, Giovannetti} (graphene with a band gap is called a gap modification of graphene).

\begin{figure}[!t]
\hypertarget{Fig2}{}
\includegraphics[width=0.5\textwidth]{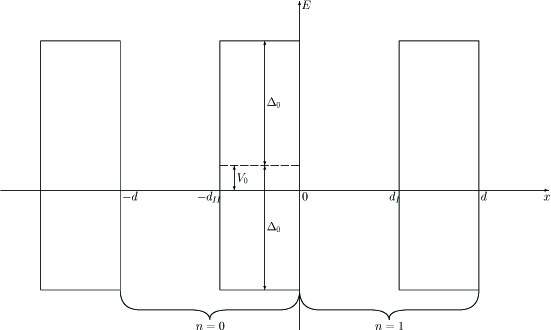}
{\bf Fig. 2.} One-dimensional periodic Kronig–Penney potential of the superlattice show n in \hyperlink{Fig1}{Fig. 1}: the periodically alternating gap modification of graphene on h-BN with a band gap of 2$\Delta_0$ = 53 meV and gapless graphene on SiO$_2$.
\end{figure}

It is assumed that all of the contacts between the regions of different band gaps are first-kind contacts (Dirac points of graphene are located in the band gaps of the gap modification of graphene). Such a superlattice is a first-type superlattice (classification of superlattices can be found, e.g., in \cite{Silin}).

There are other gap modifications of graphene. A band gap in graphene epitaxially grown on the SiC substrate is nonzero \cite{Mattausch} and, according to experiments with the use of angular resolved photoemission spectroscopy, is 0.26 eV \cite{Berkeley}. An additional modification of graphene, graphane \cite{Elias}, where the direct band gap at the point $\Gamma$ is 5.4 eV according to the calculations \cite{Lebegue}, was recently synthesized by hydration. However, the insertion of epitaxially grown graphene between gapless graphene strips is not easier than the insertion of h-BN. The band gap of graphane is too wide. These gap modifications of graphene are inapplicable for superlattices.

The main advantage of the proposed superlattice is the simplicity of the manufacture and control
of its periodicity. It is worth noting that some problems can arise in the superlattice. The difference between the lattice constants of h-BN and graphene is about~2\%~\cite{Giovannetti}. If about 100 hexagonal graphene cells are packed into one period of the superlattice, the formation of the band gap in the gap modification of graphene in the graphene sheet regions above h-BN is violated owing to the inaccurate arrangement of carbon atoms above boron or nitrogen atoms. Since contacts between graphene and its gap modification are not heterocontacts (contacts between substances with different chemical compositions), the edges of quantum wells can be insufficiently sharp and quantum wells cannot be considered as square quantum wells. A transient layer with a spatially varying band gap can exist instead of the sharp edge. Finally, the substrate can be stressed. The appearing periodic stress field of the substrate can also affect the band structure of the proposed superlattice, but this effect is very small.

\begin{center}
2. MODEL DESCRIBING THE SUPERLATTICE
\end{center}

The $x$ and $y$ axes are perpendicular and parallel to the interfaces of h-BN and SiO$_2$ strips, respectively (see \hyperlink{Fig1}{Fig. 1}). The superlattice is described by the Dirac equation
\begin{equation}\label{1}
\left(v_F{\boldsymbol\sigma}\widehat{\bf p}+\Delta\sigma_z+V\right)\Psi(x\,y)=E\Psi(x,\,y),
\end{equation}
where $v_F\approx10^8$ cm/s is the Fermi velocity, ${\boldsymbol\sigma}=(\sigma_x,\, \sigma_y)$ and $\sigma_z$ are the Pauli matrices, and $\widehat{\bf p}=-i{\boldsymbol\nabla}$ is the
momentum operator (the system of units with $\hbar$ = 1 is used). The half-width of the band gap is periodically modulated:
\begin{equation*}
\Delta=\begin{cases} 0,& d(n-1)<x<-d_{II}+dn,\\
\Delta_0,& -d_{II}+dn<x<dn,
\end{cases}
\end{equation*}
where $n$ is an integer enumerating the supercells of the superlattice; $d_I$ and $d_{II}$ are the widths of the SiO$_2$ and h-BN strips, respectively; and  $d$ = $d_I$ + $d_{II}$ are the period of the superlattice (the size of the supercell along the  x axis). The periodic scalar potential $V$ can appear due to the difference between the energy positions of the middle of the band gap of the gap modification of graphene and conic points of the Brillouin zone of gapless graphene (see \hyperlink{Fig2}{Fig. 2}):
\begin{equation*}
V=\begin{cases} 0,& d(n-1)<x<-d_{II}+dn,\\ V_0,& -d_{II}+dn<x<dn.\end{cases}
\end{equation*}
In order for the superlattice to be a first-type superlattice, the inequality $|V_0|$ $\leq$ 0 should be satisfied. The solution of Eq. \eqref{1} for the first supercell has the form
\begin{equation*}
\Psi(x, y)=\psi_1(x)e^{ik_yy}, \hspace{0.5cm} 0<x<d.
\end{equation*}
For the $n$th supercell, in view of the periodicity of the superlattice,
\begin{equation*}
\psi_n(x)=\psi_1(x+(n-1)d).
\end{equation*}
In the quantum well region (0 $<$ $x$ $<$ $d_I$), the solution of Eq. \eqref{1} is a plane wave
\begin{equation}\label{2}
\psi^{(1)}_n(x)=N_{k_1}{a^{(1)}_n\choose b^{(1)}_n}e^{ik_1x}+N_{k_1}{c^{(1)}_n\choose d^{(1)}_n}e^{-ik_1x},
\end{equation}
where $N_{k_1}$ is the normalization factor. The substitution of Eq. \eqref{2} into Eq. \eqref{1} provides the relation between the lower and upper spinor components
\begin{equation*}
b^{(1)}_n=\lambda_+a^{(1)}_n, \hspace{0.25cm} d^{(1)}_n=-\lambda_-c^{(1)}_n, \hspace{0.25cm} \lambda_\pm=\frac{v_F(k_1\pm ik_y)}{E}.
\end{equation*}
The relation of $E$ with $k_1$ and $k_y$ has the form
\begin{equation*}
E=\pm v_F\sqrt{k^2_1+k^2_y}.
\end{equation*}
It is convenient to represent Eq. \eqref{2} in a more compact form \cite{Barbier}:
\begin{equation}\label{3}
\begin{split}
\psi^{(1)}_n(x)&=\Omega_{k_1}(x){a^{(1)}_n\choose c^{(1)}_n},\\
\Omega_{k_1}(x)&=N_{k_1}\begin{pmatrix}1&1\\
\lambda_+&-\lambda_-\end{pmatrix}e^{ik_1x\sigma_z}.
\end{split}
\end{equation}
When the inequality
\begin{equation}\label{4}
\Delta^2_0+v^2_Fk^2_y-(E-V_0)^2\geq0
\end{equation}
is satisfied, the solution of Eq. \eqref{1} in the barrier region ($d_I$ $<$ $x$ $<$ $d$) has the form
\begin{equation}\label{5}
\begin{split}
\psi^{(2)}_n(x)&=\Omega_{k_2}(x){a^{(2)}_n\choose c^{(2)}_n},\\
\Omega_{k_2}(x)&=N_{k_2}\begin{pmatrix}1&1\\
-\widetilde{\lambda}_-&\widetilde{\lambda}_+\end{pmatrix}e^{k_2x\sigma_z},
\end{split}
\end{equation}
where
\begin{equation*}
\widetilde{\lambda}_\pm=\frac{iv_F(k_2\pm k_y)}{E+\Delta_0-V_0},\hspace{0.1cm}k_2=\frac{1}{v_F}\sqrt{\Delta^2_0+v^2_Fk^2_y-(E-V_0)^2}.
\end{equation*}
The solution of Eq. \eqref{1} in the barrier region under the condition
\begin{equation}\label{6}
\Delta^2_0+v^2_Fk^2_y-(E-V_0)^2<0
\end{equation}
is given by Eq. \eqref{5} with the change $k_2\rightarrow i\varkappa_2$, i.e., it is oscillating.

The possibility of existing \emph{Tamm} minibands formed by localized states near the interface between graphene and its gap modification will be considered below. In this case, $k_1\rightarrow i\varkappa_1$ $k_2$ is real. A necessary condition for existing  Tamm states has the form
\begin{equation*}
|k_y|\geq|\varkappa_1|;
\end{equation*}
under this condition, the energy $E=\pm v_F\sqrt{k^2_y-\varkappa^2_1}$ is real.

\begin{center}
3. DERIVATION OF THE DISPERSION\\ RELATION
\end{center}

The dispersion relation is derived using the transfer matrix ($T$ matrix) method. The $T$ matrix relates the spinor components for the $n$th supercell to the spinor components of the solution of the same type for the ($n$ + 1)th supercell. For example, for the solution in the quantum well region,
\begin{equation}\label{7}
{a^{(1)}_{n+1}\choose c^{(1)}_{n+1}}=T{a^{(1)}_n\choose c^{(1)}_n}.
\end{equation}
To determine the $T$ matrix, the following conditions of the continuity of the solution of the Dirac equation describing the considered superlattice are used:
\begin{equation*}
\begin{split}
\psi^{(1)}_n(d_I-0)&=\psi^{(2)}_n(d_I+0),\\
\psi^{(2)}_n(d-0)&=\psi^{(1)}_{n+1}(+0).
\end{split}
\end{equation*}
These conditions provide the equalities
\begin{equation*}
\begin{split}
{a^{(2)}_n\choose c^{(2)}_n}&=\Omega^{-1}_{k_2}(d_I)\Omega_{k_1}(d_I){a^{(1)}_n\choose c^{(1)}_n},\\
{a^{(1)}_{n+1}\choose c^{(1)}_{n+1}}&=\Omega^{-1}_{k_1}(0)\Omega_{k_2}(d){a^{(2)}_n\choose c^{(2)}_n}.
\end{split}
\end{equation*}
According to definition (7) of the T matrix and the last
two equalities\footnote{Note that the cyclic permutations of the factors of $\Omega$ matrices are possible in the definition of the $T$ matrix; these permutation do not change dispersion relation \eqref{10}. This can be verified by comparing Eq. \eqref{8} with Eq. (23) in \cite{Barbier}.},
\begin{equation}\label{8}
T=\Omega^{-1}_{k_1}(0)\Omega_{k_2}(d)\Omega^{-1}_{k_2}(d_I)\Omega_{k_1}(d_I).
\end{equation}

The substitution of Eqs. \eqref{3} and \eqref{5} with the corresponding arguments into Eq. \eqref{8} yields the expressions
\begin{equation}\label{9}
\begin{split}
T_{11}=\alpha
e^{ik_1d_I}\left[(\lambda_-+\widetilde{\lambda}_+)(\lambda_++\widetilde{\lambda}_-)e^{-k_2d_{II}}-\right.\\
\left.-(\lambda_--\widetilde{\lambda}_-)(\lambda_+-\widetilde{\lambda}_+)e^{k_2d_{II}}\right],\\
T_{12}=2\alpha
e^{-ik_1d_I}(\lambda_-+\widetilde{\lambda}_+)(\lambda_--\widetilde{\lambda}_-)sh(k_2d_{II}),\\
T_{21}=T_{12}^*,\hspace{0.25cm}T_{22}=T_{11}^*,
\end{split}
\end{equation}
where
\begin{equation*}
\alpha=\frac{1}{(\lambda_++\lambda_-)(\widetilde{\lambda}_++\widetilde{\lambda}_-)}.
\end{equation*}

The last two relations in Eqs. \eqref{9} are the general properties of the $T$ matrix.

The derivation of the dispersion relation with the use of the $T$ matrix is briefly as follows.

Let $N$ = $L/d$ be the number of supercells in the entire superlattice, where L is the length of the superlattice along the  x axis, i.e., the direction of the application of the periodic potential. The Born–Karman cyclic boundary conditions for the superlattice have the form
\begin{equation*}
\psi^{(1,2)}_N(x)=\psi^{(1,2)}_1(x).
\end{equation*}
At the same time,
\begin{equation*}
\psi^{(1,2)}_N(x)=T^N\psi^{(1,2)}_1(x),
\end{equation*}
from which, $T^N=\mathcal{I}$, where $\mathcal{I}$ is the $2\times2$ unit matrix.

It is convenient to diagonalize the $T$ matrix by means of the transition matrix $S$:
\begin{equation*}
T_d=STS^{-1}=\begin{pmatrix}\lambda_1&0\\0&\lambda_2\end{pmatrix},
\end{equation*}
where $\lambda_{1,2}$ are the eigenvalues of the $T$ matrix and have the property $\lambda_2=\lambda_1^*$. According to $T_d^N=\mathcal{I}$
\begin{equation*}
\lambda_1=e^{2\pi in/N}, \hspace{0.25cm}-N/2<n\leq N/2.
\end{equation*}

In view of the property $TrT=TrT_d$ and in terms of the notation $k_x=2\pi n/L$ ($-\pi/d<k_x\leq\pi/d$), the dispersion relation is obtained in the form
\begin{equation}\label{10}
TrT=2\cos(k_xd).
\end{equation}
Taking into account the last relation in Eqs. \eqref{9}, Eq.~\eqref{10} can also be written in the form
\begin{equation*}
ReT_{11}=\cos(k_xd).
\end{equation*}
Dispersion relation \eqref{10} under condition \eqref{4} gives the equation
\begin{equation}\label{11}
\begin{split}
&\frac{v^2_Fk^2_2-v^2_Fk^2_1+V^2_0-\Delta^2_0}{2v^2_Fk_1k_2}sh(k_2d_{II})\sin(k_1d_I)\\
&+ch(k_2d_{II})\cos(k_1d_I)=\cos(k_xd).
\end{split}
\end{equation}
According to this equation, the passage to the single-band limit is performed by two methods: first, $V_0$~=~$\Delta_0$ (the quantum well only for electrons) and, second, $V_0$ = $-\Delta_0$ (the quantum well only for holes). The result of the passage coincides with the known nonrelativistic dispersion relation (see, e.g., \cite{Herman}), although the expressions for $k_1$, $k_2$, and $E$ are different.

If inequality \eqref{6} is satisfied, the change $k_2\rightarrow i\varkappa_2$ should be made in Eq. \eqref{11}:
\begin{equation}\label{12}
\begin{split}
&\frac{-v^2_F\varkappa^2_2-v^2_Fk^2_1+V^2_0-\Delta^2_0}{2v^2_Fk_1\varkappa_2}\sin(\varkappa_2d_{II})\sin(k_1d_I)\\
&+\cos(\varkappa_2d_{II})\cos(k_1d_I)=\cos(k_xd).
\end{split}
\end{equation}
For  Tamm minibands, the change $k_1\rightarrow i\varkappa_1$ should be made in Eq. \eqref{11}:
\begin{equation}\label{13}
\begin{split}
&\frac{v^2_Fk^2_2+v^2_F\varkappa^2_1+V^2_0-\Delta^2_0}{2v^2_F\varkappa_1k_2}sh(k_2d_{II})sh(\varkappa_1d_I)\\
&+ch(k_2d_{II})ch(\varkappa_1d_I)=\cos(k_xd).
\end{split}
\end{equation}
Equation \eqref{13} has the solution only under the condition
\begin{equation*}
v^2_Fk^2_2+v^2_F\varkappa^2_1+V^2_0-\Delta^2_0<0.
\end{equation*}
However, this inequality is never satisfied; correspondingly, the \emph{Tamm} minibands are absent in the superlattice under consideration.

\begin{center}
4. RESULTS OF THE NUMERICAL\\ CALCULATION
\end{center}

The numerical calculations of the dependence of the energy on $k_x$ were performed for two values  $k_y$ = 0 and 0.1 nm$^{–1}$ at $V_0$ = 0 (see \hyperlink{Fig3}{Fig. 3}). The energy of carriers is assumed to be low, $|E|$ $\lesssim$ 1 eV, because the Dirac dispersion relation for carriers and, correspondingly, Dirac equation \eqref{1} are invalid for high energies.

The electron minibands are separated from hole minibands by a band gap, which increases with $|k_y|$. For $d_I=d_{II}$ at $k_y=0$, it is $E_g$ $\simeq$ 10–30 meV when d=10---100 nm. In this case, the solution of Eq. \eqref{11} is transformed to the solution of Eq. \eqref{12}. The band gap can increase strongly when $d_{II}$ increases with respect to $d_I$: $E_g$ $\gtrsim$ 100 meV, i.e., is several times larger than $2\Delta_0$.

The width of the minibands decreases with an increase in the period of the superlattice $d$. The dependence of the width of the minibands on $V_0$ was also examined. The widths of the electron and hole minibands increase and decrease, respectively, at $V_0$ $>$ 0 and vice versa at $V_0$ $<$ 0.

\begin{center}
\begin{figure}[!t]
\hypertarget{Fig3}{}
\includegraphics[width=0.446\textwidth]{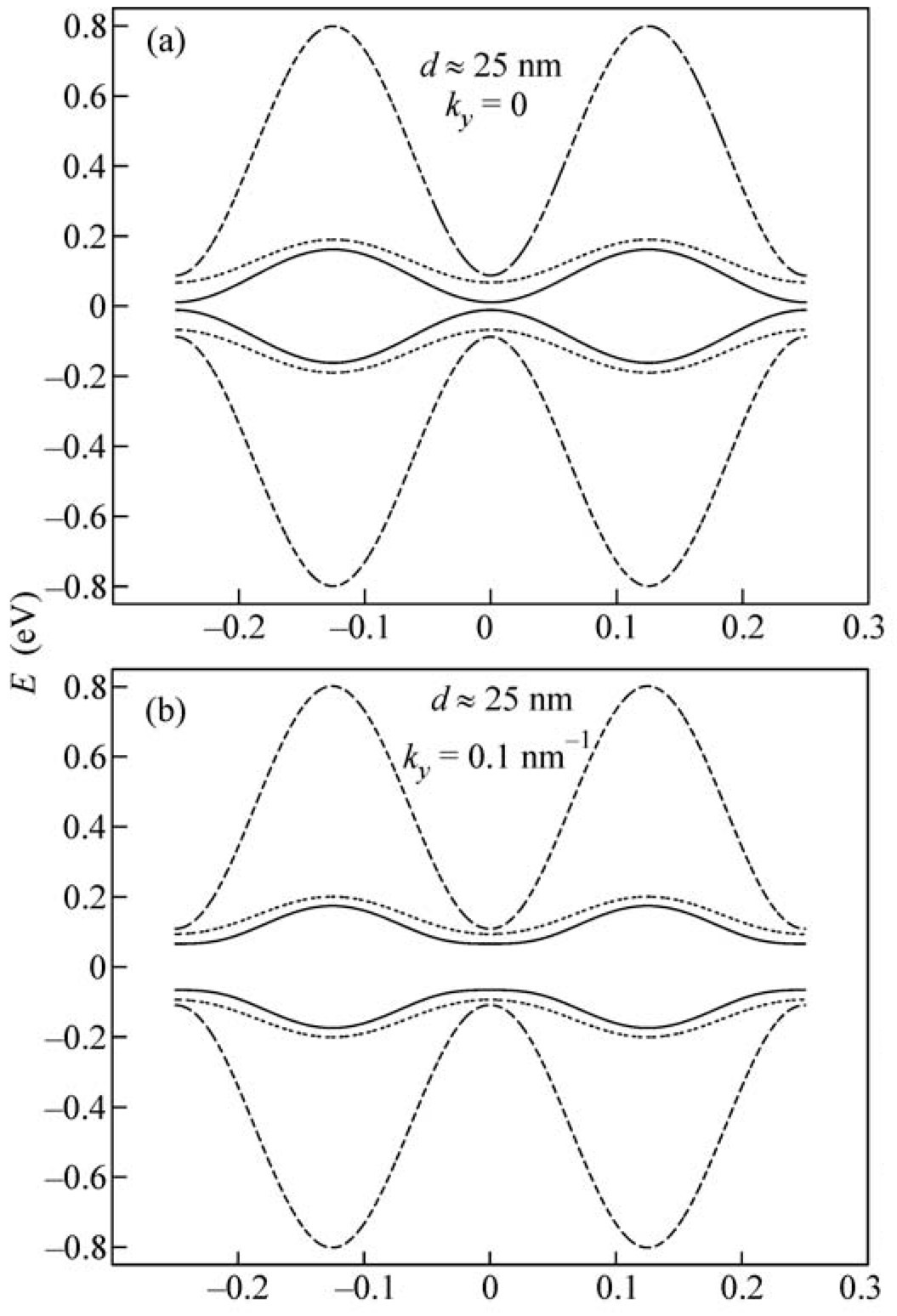}\\
\includegraphics[width=0.446\textwidth]{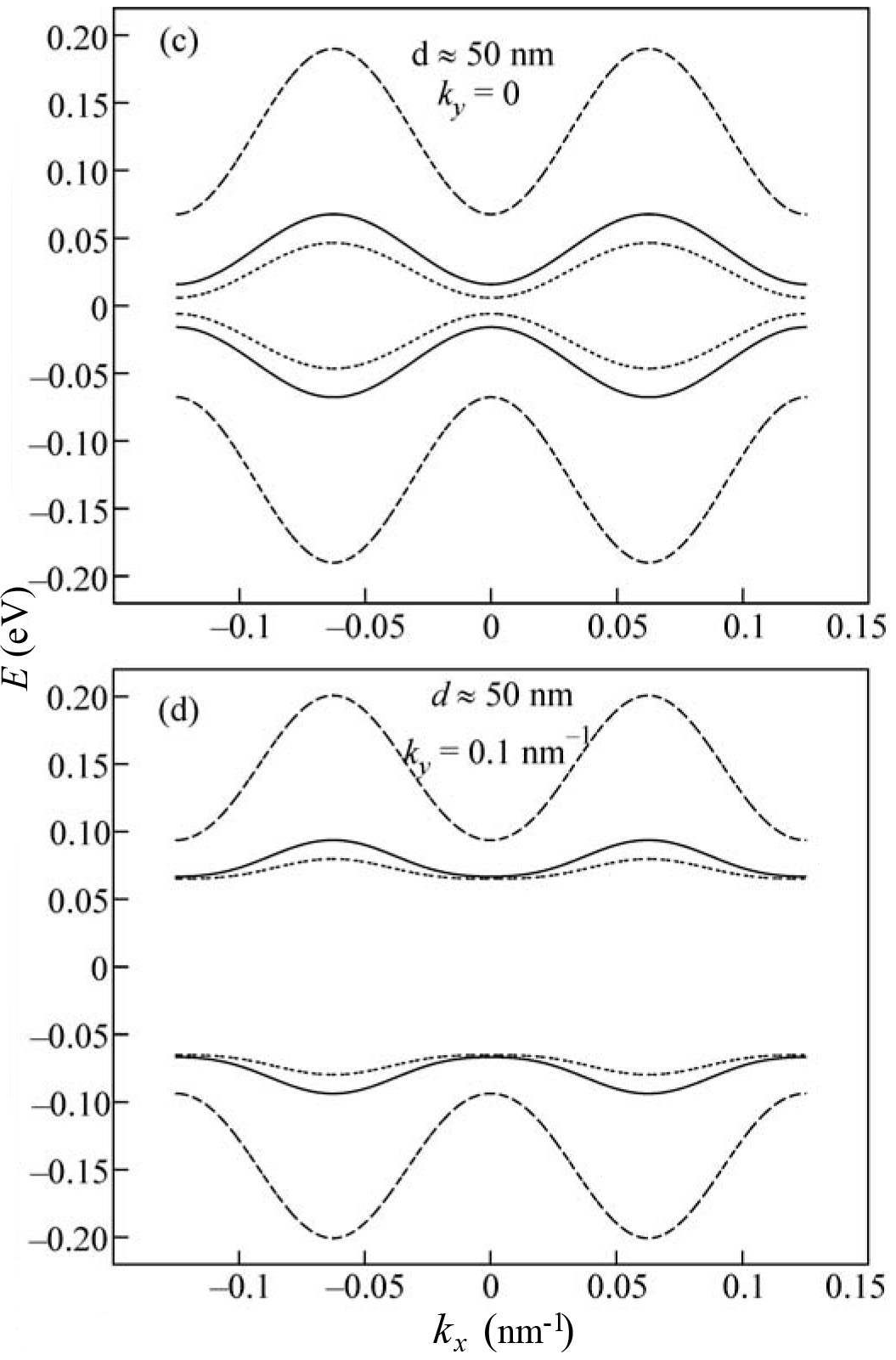}\\
{\bf Fig. 3.} Numerically calculated dependence of the energy on $k_x$ for two $k_y$ values and two superlattice periods d. The dispersion curves for the superlattices with $d_I$ = (solid lines) $d_{II}$, (dashed lines) $d_{II}/2$, and (dotted lines) $2d_{II}$.
\end{figure}
\end{center}

\newpage

\begin{center}
5. POSSIBLE APPLICATIONS\\ OF THE SUPERLATTICE
\end{center}

The described superlattice can be used as a field-effect transistor (FET) where the substrate serves as a gate. The ratio of the current through the superlattice to the current through the gate at a substrate thickness of about 10 nm can reach $\sim$ $10^6$ as for FET based on graphene nanoribbons \cite{Wang}. The main advantage of the considered superlattice is the absence of the effect of the scattering of carriers on the edges of a nanoribbon on their mobility. The mobility of the carriers in gapless graphene reaches $\mu_0$~=~2~$\times$~$10^5$~cm$^2$/(V s) \cite{Du, Morozov}. However, the mobility of carriers in FET based on the graphene nanoribbon with a width of $w$~$\sim$~3~nm is three orders of magnitude smaller than $\mu_0$. The cause of such a strong decrease is possibly the scattering of carriers at the edges of graphene nanoribbons. The mean free path between two acts of the scattering of carriers at the edge of the graphene nanoribbon $\lambda_{edge}\propto w$/P, where P is the probability of backscattering \cite{Wang}. For sufficiently good edges, P$\ll$1. The problem of scattering on edges is absent for the proposed superlattice; for this reason, the mobility of the carriers in the superlattice is expected to be $\sim\mu_0$ in the absence of the problems with the periodicity of  the potential. At the same time, a sufficiently large $E_g$ value, which provides the operation of FET at room temperature, can be reached.

If an Au film is deposited on the lower side of the substrate and graphene is optically pumped, the superlattice can be used as a terahertz laser similar to a terahertz laser based on gapless graphene \cite{Aleshkin}. In this case, terahertz radiation will be emitted from the regions of the SiO$_2$ substrate.

\vspace{0.5cm}

\begin{center}
6. CONCLUSIONS
\end{center}

A model describing such a superlattice based on graphene on a strip substrate has been proposed. The dispersion relation has been derived, which is transferred to the known nonrelativistic dispersion relation in the passage to the single-band limit. The numerical calculations have been performed for a pair of the nearest electron and hole minibands using the derived dispersion relation. Possible applications of the superlattice as a transistor or a terahertz laser have been pointed out.

\vspace{0.5cm}

\begin{center}
ACKNOWLEDGMENTS
\end{center}

I am grateful to A.P. Silin for valuable advice and stimulating discussion of the results of the work. This work was partially supported by the Dynasty Foundation; by the Educational–Scientific Complex, P.N.~Lebedev Physical Institute; and by the Presidium of the Russian Academy of Sciences (program for support of young scientists).

\end{document}